\documentclass[a4paper]{article}

\usepackage{INTERSPEECH2021}
\usepackage{marvosym}
\usepackage{multirow}
\usepackage{booktabs}
\usepackage{cite}
\usepackage{verbatim}
\usepackage{color}
\usepackage[colorlinks,linkcolor=blue,]{hyperref}

\title{Joint Echo Cancellation and Noise Suppression based on Cascaded Magnitude and Complex Mask Estimation}
\name{Xiaofeng Shu\textsuperscript{$\ast$}\thanks{$^\ast$~~Equal contribution}, Yehang Zhu\textsuperscript{$\ast$}, Yanjie Chen\textsuperscript{\Letter}\thanks{\textsuperscript{\Letter}~Corresponding author: chenyanjie@bytedance.com},  Li Chen, Haohe Liu, \\Chuanzeng Huang and Yuxuan Wang}
\address{
	Speech, Audio and Music Intelligence (SAMI) group, ByteDance}
\email{\{shuxiaofeng, zhuyehang, chenyanjie, chenli.cloud, liuhaohe.7, huangchuanzeng, wangyuxuan.11\}@bytedance.com}

\begin{document}
	
	\maketitle
	\begin{abstract}
% 		Understanding speech degraded by acoustic echo and background noise is a difficult task for human beings. 
		Acoustic echo and background noise can seriously degrade the intelligibility of speech. In practice, echo and noise suppression are usually treated as two separated tasks and can be removed with various digital signal processing (DSP) and deep learning techniques. 
		In this paper, we propose a new cascaded model, magnitude and complex temporal convolutional neural network (MC-TCN), to jointly perform acoustic echo cancellation and noise suppression with the help of adaptive filters. The MC-TCN cascades two separation cores, which are used to extract robust magnitude spectra feature and to enhance magnitude and phase simultaneously. 
% 		In this paper, a joint method based on the is proposed to suppress both residual echo and background noise in the output of the adaptive filter. 
		Experimental results reveal that the proposed method can achieve superior performance by removing both echo and noise in real-time. In terms of DECMOS, the subjective test shows our method achieves a mean score of 4.41 and outperforms the INTERSPEECH2021 AEC-Challenge baseline by 0.54.
	\end{abstract}
	
	\section{Introduction}
	Joint acoustic echo cancellation (AEC) and noise suppression (NS) aim at improving the quality and intelligibility of speech corrupted by environmental noises and echos, which can affect many real-world communication applications. Currently, a large numbers of AEC and NS systems have been proposed based on digital signal processing techniques~\cite{loizou2013speech, cohen2001speech, deb2014technical, turbin1997comparison}. Traditional noise suppression algorithms are introduced in detail in \cite{loizou2013speech}, including spectral subtraction, Wiener filtering, and some statistical-model-based methods. In \cite{cohen2001speech}, An optimally-modified log-spectral amplitude (OM-LSA) speech estimator was presented, which can effectively reduce the stationary noises. In the AEC task, adaptive filtering is frequently used to filter the far-end signal in order to simulate the echo path and obtain an estimation of acoustic echo. \cite{deb2014technical} discussed several adaptive algorithms for echo cancellation system. In \cite{turbin1997comparison}, a residual echo suppressor was used to further attenuate echos. However, these methods are unable to remove highly non-stationary noise and residual echo in real-world cases.
	%Although they are effective in certain situations, they may also add extra distortions to the near-end signal and even degrade clean speech in some circumstances.
	
	\begin{figure}[t]
		\begin{center}
			\includegraphics[width=0.45\textwidth]{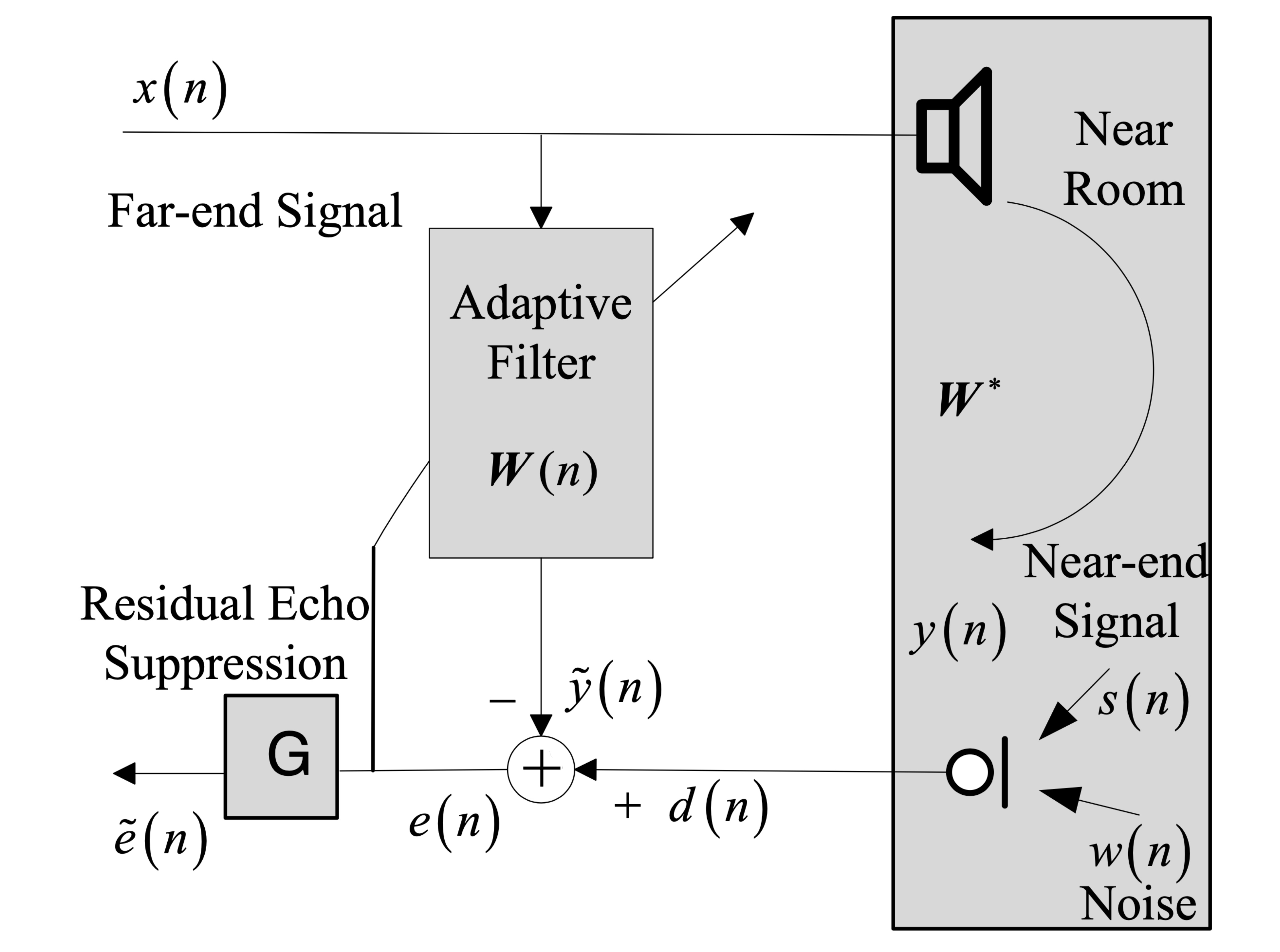}
		\end{center}
		\caption{
			A typical acoustic echo cancellation system.}
		\label{fig:aec_system}
	\end{figure}
	
	Deep neural network (DNN)-based architectures have been successfully applied to noise suppression~\cite{chen2017long, zhang2020deepmmse, hu2020dccrn, tan2018convolutional, westhausen2020dual}, which greatly motivates the investigations of DNN for AEC~\cite{lee2015dnn, zhang2018deep}. One of the most popular DNN structures is the long short-term memory (LSTM), which has been applied to noise suppression task~\cite{chen2017long} due to its capability of temporal modeling. A temporal convolutional network (TCN) is designed for noise suppression task~\cite{zhang2020deepmmse} and outperformed the residual long short-term memory (ResLSTM) network \cite{nicolson2019deep} with fewer parameters. Deep complex convolution recurrent network (DCCRN)~\cite{hu2020dccrn}, which outperforms CRN~\cite{tan2018convolutional} by a large margin, estimates complex ratio mask (CRM) and optimizes using signal approximation(SA). In ~\cite{westhausen2020dual}, the authors proposed a cascaded model structure for noise suppression, which demonstrates the advantage of using two types of analysis and synthesis bases in a stacked network approach. In ~\cite{lee2015dnn}, a deep neural network (DNN)-based residual echo suppression (RES) gain estimation in all frequency bins was proposed based on both the far-end and the AEC output signals.
	In~\cite{zhang2018deep}, bidirectional long short-term memory (BLSTM) is trained to estimate the ideal ratio mask from features extracted from the mixtures of near-end and far-end signals. 
	The methods mentioned above focus solely on echo or noise suppression. % todo
	However, in the real-world scenario, echos and noises usually exist simultaneously. In this case, deep learning-based echo cancellation and noise suppression methods demonstrated strong capabilities and emerged as state-of-the-art solutions. Furthermore, in ~\cite{zhang2019deep},  a convolutional recurrent network (CRN) was utilized to estimate the real and imaginary spectrograms of near-end speech. To improve the main task of estimating the near-end signal, ~\cite{fazel2019deep} proposed a multi-task learning scheme based on deep gated recurrent neural networks to learn the auxiliary task, i.e., estimating the echo. To ensure robust and effective information processing, ~\cite{westhausen2020acoustic} suggested using a dual-signal transformation LSTM network (DTLN) that incorporates a short-time Fourier transformation and a learned feature representation.
	
	In this paper, motivated by the TCN-based speech enhancement~\cite{zhang2020deepmmse} and  DTLN-based AEC~\cite{westhausen2020acoustic}, we propose a new cascaded network, termed as magnitude and complex temporal convolutional neural network (MC-TCN), to jointly perform AEC and NS. The proposed cascaded model consists of two separation cores. The first core uses real-value units to predict the magnitude mask, with which we can obtain a robust magnitude estimation.
	%as the learned feature representation. 
	Then we combine this estimation with the phase of the input signal to construct the input to the second separation core. The second core transforms the input into a complex spectrogram and uses complex-valued units to estimate CRM to further enhance magnitude and phase. By employing TCN and complex TCN as the separation cores estimators, our approach is easy to perform parallelization and it is more light-weighted comparing with models like LSTM and complex LSTM. The experiment results demonstrate that our proposed model can effectively suppress acoustic echo and noise in both single-talk and double-talk scenarios.
	
	The rest of the paper is organized as follows. Section~\ref{sec:method} describes the joint echo cancellation and noise suppression system as well as their sub-modules. In Section~\ref{sec:exp}, the experimental and comparison results are presented to assess the performance of the proposed method. The conclusions are finally given in Section~\ref{sec:conclusion}.
	
	\section{Proposed method}
	\label{sec:method}
	% In this section, we include a brief description and detailed introduction to the system and its modules in this study.
	\subsection{Signal model}
	The signal model for a joint echo cancellation and noise suppression system is illustrated in Figure~\ref{fig:aec_system}. The near-end microphone signal is given by
	\begin{equation}
		d(n) = s(n) + y(n) + w(n),
	\end{equation}
	where $s(n)$ is clean speech and $w(n)$ is environmental noise, and the acoustic echo signal can be expressed as 
	\begin{equation}
	    {y}(n) = W^{*} \star x(n),
	\end{equation}
	 in which $W^{*}$ is the room impulse response (RIR) and $\star$ denotes the convolution. The error signal $e(n)$ can be obtained by subtracting the estimated echo $\tilde{y}(n)$ from near-end signal $d(n)$. Since the estimation $\tilde{y}(n)$ is not guaranteed to be perfect, $e(n)$ may still contains residual echo and environmental noise. 
% 	 To further attenuate echo and noise in $e(n)$, a joint residual echo suppression (RES) and NS method, MC-TCN, is used.
	
	\subsection{System overview}
	\label{sec:system}
	\begin{figure}[ht]
		\begin{center}
			
			\includegraphics[width=0.5\textwidth]{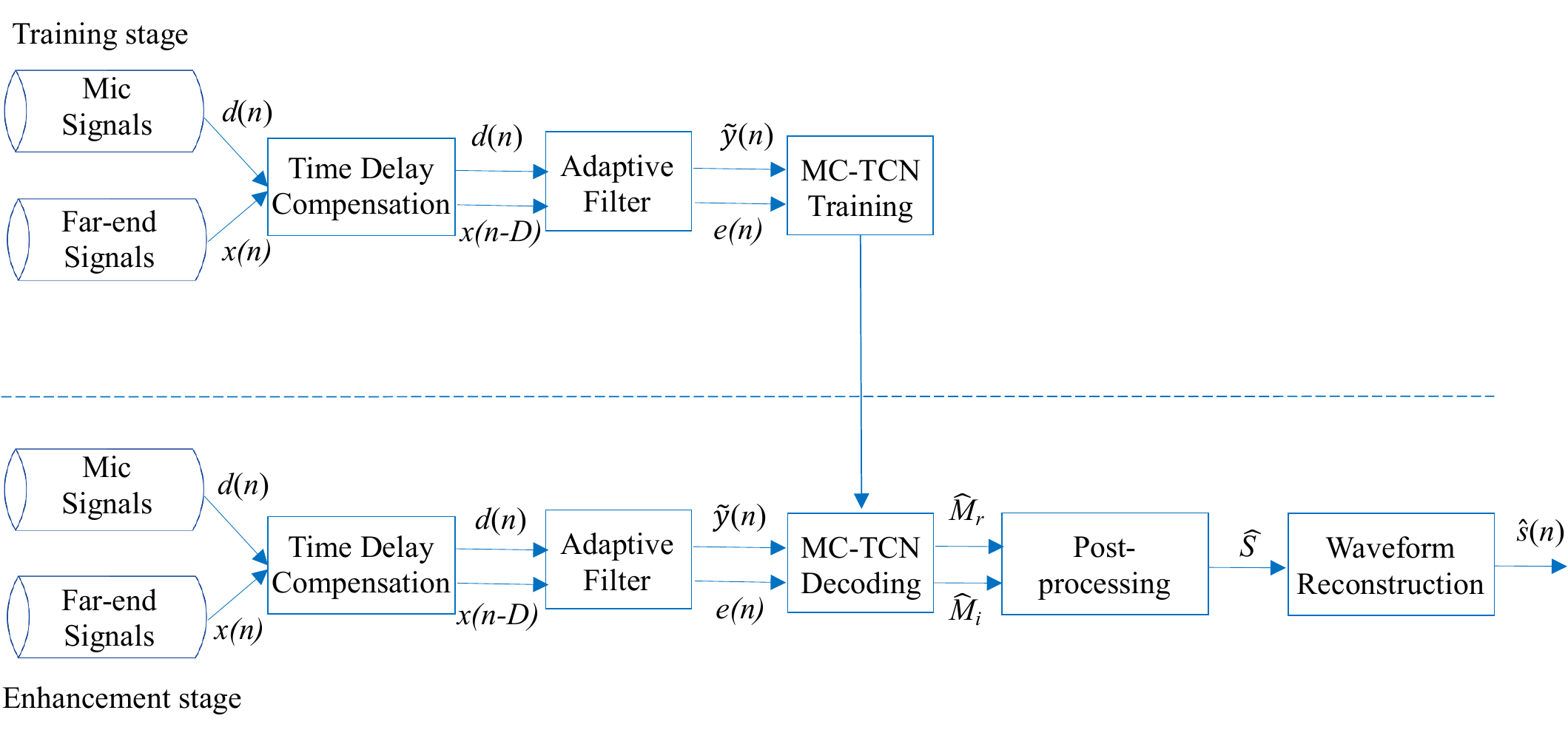}
			
		\end{center}
		\caption{
			Schematic diagram of the MC-TCN based joint AEC and NS. }
		\label{fig:overview}
	\end{figure}
	
	To further attenuate echo and noise in $e(n)$, we design a joint RES and NS method. The schematic diagram of our MC-TCN based system is illustrated in Figure~\ref{fig:overview}. The whole system consists of two stages. First, in the ``training stage", a time delay compensation algorithm is applied to estimate the delay between the far-end signal $x(n)$ and the near-end signal $d(n)$. 
	If the estimated delay $D$ is too large, the far-end signal will be delayed for compensation. After that, adaptive filtering is performed to suppress the linear part of acoustic echo in $d(n)$.
	The estimated echo ${\tilde{y}}(n)$ and the error signal $e(n)$ will be used as two inputs of the neural network. Here, $e(n)$ contains clean speech $s(n)$, background noise $w(n)$ and residual echo $y(n)-{\tilde{y}}(n)$. % todo coherence & fomula
	%Finally, MC-TCN is trained using $e(n)$, ${\tilde{y}}(n)$ and $s(n)$.
	Finally, $e(n)$ and ${\tilde{y}}(n)$ are fed into MC-TCN to predict a complex ratio mask for the enhancement of $e(n)$, and $s(n)$ is our training target during optimization. In our experiments, we found that using $e(n)$ and ${\tilde{y}}(n)$ as inputs outperforms using $e(n)$ and  aligned $x(n)$.
	
	In the ``enhancement stage", the same as training stage, we first perform time delay compensation and adaptive filtering on $d(n)$ and $x(n)$ to obtain ${\tilde{y}}(n)$ and $e(n)$. The well-trained MC-TCN then takes ${\tilde{y}}(n)$ and $e(n)$ as input to predict the real mask $\hat{M}_r$ and imaginary mask $\hat{M}_i$. A post-processing module is used to calculate the estimated near-end clean speech spectrum $\hat{S}$. After the reconstructed spectrum is obtained, overlap-add~\cite{du2008speech} is used to reconstruct the waveform of the estimated near-end clean speech $\hat{s}(n)$.
	
	\subsection{Time delay compensation}
	\label{sec:timedelay}
	There is always an unknown delay between far-end signal $x(n)$ played by loudspeaker and near-end signal $d(n)$ captured by microphone due to hardware-related latency and software-related buffering mechanism. We employ an algorithm based on audio fingerprint~\cite{cano2005review} to estimate possible delay $D$ within 1000 ms between $x(n)$ and $d(n)$. We use spectral peaks as our features and combine them together to form a pattern. The pattern and the calculated delay time was updated every 16ms.
	When the detected delay is longer than 250 ms, we will use the delayed version of far-end signal $x(n-D)$ as the reference signal of the adaptive filter.
	
	\subsection{Adaptive filter}
	\label{sec:adaptivefilter}
	The adaptive filter used in our system is derived from speexDSP's multi delay block frequency-domain (MDF) adaptive filter algorithm~\cite{soo1990multidelay}. The divergence of the adaptive filter is controlled with a two-echo-path model as described in~\cite{ochiai1977echo}. We also utilize learning rate control~\cite{valin2007adjusting} to increase robustness in double-talk scenarios and a block variant of the PNLMS algorithm~\cite{duttweiler2000proportionate} to speed up adaptation.
	As a compromise between complexity and convergence, a variant of AUMDF~\cite{soo1990multidelay} is used, where the block with the highest energy is constrained in each iteration and other blocks are alternatively constrained. 
	To avoid extra delay, the block size for our adaptive filter is set to 16 ms, which matches the frame size in the time delay compensation module and the input size of our MC-TCN model. We used 18 blocks during adaptive filtering which means a delay time less than 288 ms is allowed between far-end and near-end signals.
	
	\subsection{The MC-TCN model}
	\label{sec:mctcn}
	\begin{figure*}[t]
		\centering{\includegraphics[scale=0.5]{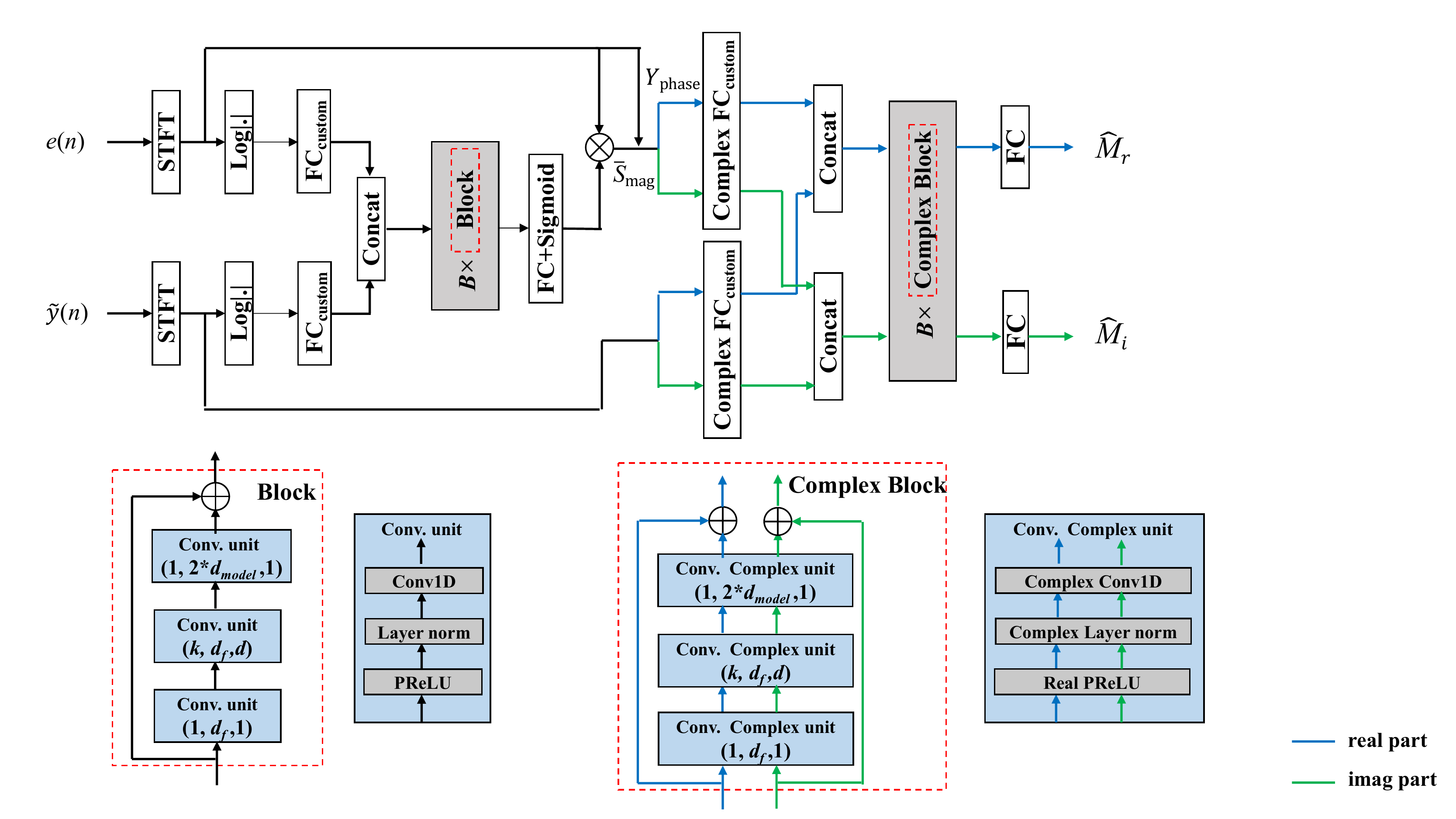}}
		\caption{
			Overview of the MC-TCN architecture. We use the adaptive filtering output $e(n)$ and the estimated echo ${\tilde{y}}(n)$ to compute the complex ratio mask $\hat{M}_r$ and $\hat{M}_i$.}
		\label{fig:MC-TCN_structure}
	\end{figure*}
	
	The MC-TCN architecture, shown in Figure~\ref{fig:MC-TCN_structure}, has two separation cores. The first separation core uses log spectra of error signal $e(n)$ and the estimated echo ${\tilde{y}}(n)$ as the input features. They are then processed by a $\rm\mathbf{FC}_{\mathbf{custom}}$ layer, which is a fully-connected layer of size ${d}_{{model}}$, including layer normalization and a parametric rectified linear unit (PReLU) activation function. The two $\rm\mathbf{FC}_{\mathbf{custom}}$ outputs are then concatenated as the input of $\mathit\mathbf{B}$ bottleneck residual blocks. As stated in~\cite{zhang2020deepmmse} and~\cite{kalchbrenner2016neural},  each block contains three one-dimensional causal dilated convolutional units. In our model, each convolutional unit is pre-activated by the PReLU activation followed by layer normalization.
	In Figure~\ref{fig:MC-TCN_structure},  we denote $\mathbf{(kernel\  size, output\  size, dilation\  rate)}$ for each convolutional unit. The output of the first separation core, which is predicted by a fully-connected layer with sigmoidal units, is multiplied by the magnitude of the linear AEC output and then transformed into a complex spectrogram using the phase of the adaptive filtering output.
	
	As for the second separation core, in Figure~\ref{fig:MC-TCN_structure}, we use blue and green lines to represent the real and imaginary parts. Similar to the processing of the first separation core, the second separation core uses similar but complex-valued operations. The complex spectrograms of the first separation core output and the estimated echo are first transformed by a $ \rm{\mathbf{Complex}\ \mathbf{FC}_{\mathbf{custom}}}$ layer, which is a fully-connected complex layer of size ${d}_{{model}}$ that includes complex layer normalization and a real-valued PReLU activation function. Then the concatenations of real parts and imaginary parts are fed into $\mathit\mathbf{B}$ bottleneck residual complex blocks. Each complex block contains three corresponding complex convolutional units and each unit is pre-activated by the real PReLU activation followed by a complex layer normalization. Finally, we use two fully connected layers to estimate the mask of the real and imaginary parts respectively.
	
	The loss function of MC-TCN during training is scale-invariant source-to-noise ratio (SI-SNR), which is calculated in time domain. The definition of SI-SNR loss is 
	\begin{equation}
		\left\{ {\begin{array}{l}
				{{s_{target}} = \frac{{\left\langle {\hat s,s} \right\rangle  \cdot s}}{{\left\| s \right\|_2^2}}}\\
				{{e_{noise}} = \hat s - s}\\
				{SI - SNR = 10*{{\log }_{10}}\left( {\frac{{\left\| {{s_{target}}} \right\|_2^2}}{{\left\| {{e_{noise}}} \right\|_2^2}}} \right)}.
		\end{array}} \right.
	\end{equation}
	The model is
	trained with the Adam optimizer~\cite{kingma2014adam} for 200 epochs with an initial
	learning rate of 5e-4. We use early stopping to select the best models and prevent overfitting.
	\subsection{Post-processing}
	As the calculation of estimated clean speech spectrum proposed in~\cite{hu2020dccrn}, we use the same method to obtain the near-end clean speech spectrum $\hat{S}$  as follows,
	\begin{equation}
		\left\{ {\begin{array}{l}
				{{{\hat M}_{{\rm{mag}}}} = \sqrt {\hat M_r^2 + \hat M_i^2} }\\
				{{{\hat M}_{{\rm{phase}}}}  = \arctan 2({{\hat M}_i},{{\hat M}_r})}\\
				{\hat S = {{\bar S}_{{\rm{mag}}}} \cdot \mathit{Tanh} ({{\hat M}_{{\rm{mag}}}}) \cdot {e^{{Y_{{\rm{phase}}}} + {{\hat M}_{{\rm{phase}}}}}}},
		\end{array}} \right.
	\end{equation}
	where ${{\bar S}_{{\rm{mag}}}} $ represents the output magnitude spectrum of the first separation core and $Y_{{\rm{phase}}}$ is the phase of error signal $e(n)$.

	\section{Experimental results}
	\label{sec:exp}
	\subsection{Datasets and experimental setup}
	INTERSPEECH2021 AEC Challenge training datasets contain synthesized data and real recordings. In this dataset, only far-end signals
	and echo signals were used as the training data. For the near-end clean speech, we chose another two datasets. One is ICASSP2021 DNS Challenge~\cite{reddy2020icassp} datasets, which contains English speech data and another is AISHELL-2~\cite{du2018aishell}. The same as~\cite{westhausen2020acoustic}, we use speech estimated by the speech enhancement method to calculate the estimate SNR. The speech file was discarded if the SNR is lower than  $5 \ \rm{dB}$. To remove echo and noise simultaneously, noise datasets in the ICASSP2021 DNS Challenge were also used as the noisy training data. To make full use of these data, all training samples were created randomly and concurrently during training. The length of the training sample is 10 seconds. 
	
	During data random selection, We assign different probability on different datasets. Specifically, to form clean speech data, we randomly chose a clean sample from DNS with $70$ percent probability, and with  $30$ percent probability, we sampled data from AISHELL-2. We adjusted the audio length to 10 seconds by cutting from a random starting point or padding zero at the end. The noise sample is also randomly selected and the audio length is altered by cutting from a random starting point or making duplication of the previous noise segment. Likewise, the length of the randomly selected echo signal and the corresponding far-end signal are both adjusted with the same random cutting point or appended with zero at the end. 
	
	To obtain the final mic signal and far-end signal in the training, we perform online augmentation with the selected data by the following four steps: 
	
	$1.$ Convolve near-end speech with different RIRs randomly-selected from the DNS RIR-dataset
	and mix noise at a specific level of SNR randomly selected from 5 dB to 25 dB.
	
	$2.$  Mix echo signal and the processed near-end speech in step $1$ with an SNR uniformly sampled from -5 dB to 20 dB.
	
	$3.$  Silent the noise sample with $20$ percent probability.
	
	$4.$  Silent the echo sample and the corresponding far-end sample with $20$ percent probability.
	
	In our work, all the waveforms are resampled to 16 kHz. During the calculation of the Short-time Fourier transform (STFT), we use a Hanning window with a frame length of 512 samples and a frame shift of 256 samples. The second convolutional unit of each block and complex block employs a dilation rate of $d$, providing a larger contextual field on the temporal dimension. % todo
	As mentioned in~\cite{luo2019conv}, the dilation rate d is cycled as the block index b increases,
	\begin{equation}
	    d = 2^{(b-1 \ \rm{mod} \ log_{2}(\mathit{D})+1)},
	\end{equation}
	in which $\rm{mod}$ is the modulo operation, and $D$ is the maximum dilation rate. In INTERSPEECH2021 AEC Challenge, we choose $d_{model}=256$, $d_{f}=64$, $k=3$ and $B=20$ as the final configuration. Finally, for each epoch, we use 300 hours of training samples created online to train the MC-TCN model.
	
	\subsection{Performance Evaluation}
	The results of the proposed joint AEC and NS method are presented in Table~\ref{tab:decmos_result},  Table~\ref{tab:diff_config} and Table~\ref{tab:aec_challenge}. To evaluate the joint AEC and NS system, we usually adopt two metrics, which are the perceptual evaluation of speech quality (PESQ) and echo return loss enhancement (ERLE). However, they are revealed to correlate poorly with human rating when used for noise suppression or acoustic echo cancellation tasks, which involves perceptually invariant transformations. Besides, these metrics are intrusive, which means that they cannot be used to evaluate real recordings. Therefore, AEC Challenge Organizer open the DESMOS API to obtain a robust objective perceptual speech quality metric. Table~\ref{tab:decmos_result} shows the DECMOS results of different systems on the blind test set of AEC Challenge. The baseline system of the AEC Challenge and DTLN-AEC system are also presented for comparisons.
	
	\begin{table}[ht]\scriptsize
		\setlength{\abovecaptionskip}{0pt}
		\centering
		\caption{DECMOS on the blindtest set of AEC Challenge.}
		\begin{tabular}{c|c|c|c|c}
			\toprule
			\multirow{1}[7]{*}[10pt]{\textbf{Type}}  &
			\multicolumn{1}{c}{\textbf{Near-end}}  & \multicolumn{1}{c}{\textbf{Far-end}}  & \multicolumn{1}{c}{\textbf{Doubletalk}} &
			\multicolumn{1}{c}{\textbf{Doubletalk}}\\
			\midrule
			\textbf{System} & \textbf{Deg MOS} & \textbf{Echo MOS} & \textbf{Echo MOS} & \textbf{Deg MOS} \\
			Baseline & 3.722 & 3.582 & 3.829  & 2.924 \\
			DTLN &3.748 & 4.076 & 3.830  & 3.264 \\
			MC-TCN &3.807 & 4.213 & 3.968  & 3.286 \\
			\bottomrule
		\end{tabular}%
		\label{tab:decmos_result}%
	\end{table}%
	
	\begin{table}[ht]\scriptsize
		\setlength{\abovecaptionskip}{0pt}
		\vspace{0.2cm}
		\centering
		\caption{the performance of MC-TCN using different block nums $B$ configuration.}
		\begin{tabular}{c|c|c|c|c}
			\toprule
			\multirow{1}[7]{*}[10pt]{\textbf{MC-TCN}}  &
			\multicolumn{1}{c}{\textbf{Near-end}}  & \multicolumn{1}{c}{\textbf{Far-end}}  & \multicolumn{1}{c}{\textbf{Doubletalk}} &
			\multicolumn{1}{c}{\textbf{Doubletalk}}\\
			\midrule
			\textbf{Block num} & \textbf{Deg MOS} & \textbf{Echo MOS} & \textbf{Echo MOS} & \textbf{Deg MOS} \\
			10 & 3.741 & 4.105 & 3.868  & 3.234 \\
			15 &3.771 & 4.249 & 3.993 & 3.247 \\
			20 &3.807 & 4.213 & 3.968  & 3.286 \\
			\bottomrule
		\end{tabular}%
		\label{tab:diff_config}%
	\end{table}%
	Table~\ref{tab:diff_config}	shows the DECMOS results for different block num $B$ configuration. Although the MC-TCN using $B=15$ suppress more echos, the distortion of the enhanced speech is worse than the one using $B=20$. We handed the enhanced results using $B=20$ into the INTERSPEECH2021 AEC-Challenge. There are totally 5.5M trainable parameters in this submitted model. The average time it takes to infer one frame is 0.7460ms on macbook pro 1.7GHz Quad-Core Core i7.
    
	\begin{table}[ht]\scriptsize
		\setlength{\abovecaptionskip}{0pt}
		\centering
		\caption{Subjective ratings in terms of MOS for the blind test set
			of the INTERSPEECH2021 AEC-Challenge (ST = single talk, DT = double talk, NE = near-end,
			FE = far-end).}
		\begin{tabular}{cccccc}
			\toprule
			\multirow{1}[7]{*}[10pt]{\textbf{Team Id}}  &
			\multicolumn{1}{c}{\textbf{ST-NE}}  & \multicolumn{1}{c}{\textbf{ST-FE}}  & \multicolumn{1}{c}{\textbf{DT-Echo}} &
			\multicolumn{1}{c}{\textbf{DT-other}} &
			\multicolumn{1}{c}{\textbf{Overall}}\\
			\midrule
			4 & 4.25 & 4.59 & 4.69  & 4.18 & 4.43\\
			2 &4.27 & 4.49 & 4.52 & 4.39 & 4.42\\
			7 &4.10& 4.54 & 4.77  & 4.24 & 4.41\\
			Ours &4.32& 4.45 & 4.59  & 4.28 & 4.41\\
			14 &4.19& 4.49 & 4.58  & 4.27 & 4.38\\
			Baseline &4.18& 3.82 & 4.04  & 3.45 & 3.87\\
			\bottomrule
		\end{tabular}%
		\label{tab:aec_challenge}%
				\vspace{0.2cm}
	\end{table}%
	    
	The top 5 results released by the organizer are shown in Table ~\ref{tab:aec_challenge}, and for overall performance, we obtain the same score with Team $7$ and ranked the third. We got a high score, especially for the ST-NE and DT-other. 
 ST-NE and DT-other scores mainly reflect the audio quality of enhanced near-end speech, which are influenced by speech distortion and residual noises, while ST-FE and DT-Echo scores reflect the performance of echo cancellation. Therefore, for the future direction of this work, the adaptive filter module will be updated by using other adaptive filter algorithms to improve the performance of echo cancellation.
	
	\section{Conclusion}
	\label{sec:conclusion}
% 	In this paper, a new joint AEC and NS method based on MC-TCN is proposed to improve the quality and intelligibility of speech corrupted by acoustic echo and background noise. 
% 	The proposed cascaded model uses two separation cores to predict magnitude mask and complex mask, which can help obtain the learned feature representation and then simultaneously enhance magnitude and phase, respectively. Magnitude TCN and complex TCN are adopted as the estimators of the separation cores. The results show that the proposed model can successfully remove echo and noise simultaneously in real-time. On the blind test-set of the AEC-Challenge, our MC-TCN based method obtains
% 	a superior performance in both single-talk and double-talk cases.
	
	In this paper, a new joint AEC and NS method based on MC-TCN is proposed to improve the quality and intelligibility of speech corrupted by acoustic echo and background noise. 
	The proposed cascaded model uses two separation cores, magnitude TCN and complex TCN, to predict magnitude masks and complex masks. These two masks are helpful to obtain the learned feature representation to simultaneously enhance magnitude and phase. The results show that the proposed model can successfully remove echo and noise simultaneously in real-time. On the blind test-set of the AEC-Challenge, our MC-TCN based method obtains a superior performance in both single-talk and double-talk cases. 
	
	% todo length in pages
	
	\bibliographystyle{IEEEtran}
	
	\bibliography{cited}
	
	% \begin{thebibliography}{9}
	% \bibitem[1]{Davis80-COP}
	%   S.\ B.\ Davis and P.\ Mermelstein,
	%   ``Comparison of parametric representation for monosyllabic word recognition in continuously spoken sentences,''
	%   \textit{IEEE Transactions on Acoustics, Speech and Signal Processing}, vol.~28, no.~4, pp.~357--366, 1980.
	% \bibitem[2]{Rabiner89-ATO}
	%   L.\ R.\ Rabiner,
	%   ``A tutorial on hidden Markov models and selected applications in speech recognition,''
	%   \textit{Proceedings of the IEEE}, vol.~77, no.~2, pp.~257-286, 1989.
	% \bibitem[3]{Hastie09-TEO}
	%   T.\ Hastie, R.\ Tibshirani, and J.\ Friedman,
	%   \textit{The Elements of Statistical Learning -- Data Mining, Inference, and Prediction}.
	%   New York: Springer, 2009.
	% \bibitem[4]{YourName17-XXX}
	%   F.\ Lastname1, F.\ Lastname2, and F.\ Lastname3,
	%   ``Title of your INTERSPEECH 2021 publication,''
	%   in \textit{Interspeech 2021 -- 20\textsuperscript{th} Annual Conference of the International Speech Communication Association, September 15-19, Graz, Austria, Proceedings, Proceedings}, 2020, pp.~100--104.
	% \end{thebibliography}
	
\end{document}